\documentclass[10pt,preprint]{aastex}
\usepackage{apjfonts,emulateapj5,psfig}

\def\ltsima{$\; \buildrel < \over \sim \;$}
\def\lsim{\lower.5ex\hbox{\ltsima}}
\def\gtsima{$\; \buildrel > \over \sim \;$}
\def\gsim{\lower.5ex\hbox{\gtsima}}

\slugcomment{}

\shorttitle{Variability of lens PMN~J1838--3427}
\shortauthors{Winn et al.}

\begin{document}

\title{The Radio Variability of the Gravitational Lens PMN~J1838--3427}

\author{
Joshua N.\ Winn\altaffilmark{1,2},
James E.J.\ Lovell\altaffilmark{3},
Hayley Bignall\altaffilmark{4},
Bryan M.\ Gaensler\altaffilmark{1},\\
Tracy J.\ Getts\altaffilmark{3},
Lucyna Kedziora-Chudczer\altaffilmark{3},
Roopesh Ojha\altaffilmark{3},
John E.\ Reynolds\altaffilmark{3},\\
Steven J.\ Tingay\altaffilmark{5},
Tasso Tzioumis\altaffilmark{3},
Mark Wieringa\altaffilmark{3}
}

\altaffiltext{1}{Harvard-Smithsonian Center for Astrophysics, 60
Garden St., Cambridge, MA 02138}

\altaffiltext{2}{National Science Foundation Astronomy \& Astrophysics
Postdoctoral Fellow}

\altaffiltext{3}{Australia Telescope National Facility, CSIRO, P.O.\
Box 76, Epping, NSW 1710, Australia}

\altaffiltext{4}{Joint Institution for VLBI in Europe, Post Bus 2,
7990 AA, Dwingeloo, The Netherlands}

\altaffiltext{5}{Centre for Astrophysics and Supercomputing, Swinburne
University of Technology, Mail No.\ 31, P.O.\ Box 218, Hawthorn,
Victoria 3122, Australia}

\begin{abstract}
We present the results of a radio variability study of the
gravitational lens PMN~J1838--3427. Our motivation was to determine
the Hubble constant by measuring the time delay between variations of
the two quasar images. We monitored the system for 4 months
(approximately 5 times longer than the expected delay) using the
Australia Telescope Compact Array at 9~GHz. Although both images were
variable on a time scale of a few days, no correlated intrinsic
variability could be identified, and therefore no time delay could be
measured. Notably, the fractional variation of the fainter image (8\%)
was greater than that of the brighter image (4\%), whereas lensed
images of a point source would have the same fractional
variation. This effect can be explained, at least in part, as the
refractive scintillation of both images due to the turbulent
interstellar medium of the Galaxy.
\end{abstract}


\section{Introduction}
\label{sec:intro}

Multiple-image gravitational lenses offer a famous and refreshingly
direct method for measuring the Hubble constant ($H_0$), as first
envisioned by Refsdal (1964, 1966). If the background source is
variable, then the variations of each image are seen at different
times, because of the different proper lengths of the image paths. A
model of the gravitational potential predicts the time delays in units
of $H_0^{-1}$. Thus, if the model is accurate, measuring the time
delays amounts to measuring the Hubble constant. Currently it is
believed that $H_0$ is known within 10\%, thanks mainly to hard-won
and high-quality data from the {\it Hubble Space Telescope}\, Key
Project (Freedman et al.\ 2001), but it is still desirable to develop
independent methods of determining $H_0$ because of the fundamental
importance of this quantity in interpreting cosmological observations.

For the time-delay method to succeed, one must obtain light curves of
multiple images that have a sufficient duration, sampling rate, and
precision for correlated intrinsic variations to be detected. This has
proven to be difficult. Time delays have been measured in only 10
systems, and many of those measurements are subject to large
uncertainties (for recent reviews, see Kochanek \& Schechter 2003 and
Courbin, Saha, \& Schechter 2002). In addition, the method is limited
by systematic errors in lens models, although at least the nature of
the degeneracy between the mass distribution and $H_0$ is well
understood (Gorenstein, Shapiro, \& Falco 1988; Kochanek 2002). This
means that if progress in measuring $H_0$ by other means continues to
outstrip progress in understanding the mass distributions of galaxies,
then the time-delay method can be run in reverse: time delays can be
used to study galaxy structure, for an assumed value of $H_0$ (see,
e.g., Kochanek 2003).

In this paper we report the results of a campaign with the Australia
Telescope Compact Array (ATCA\footnote{The ATCA is part of the
Australia Telescope, which is funded by the Commonwealth of Australia
for operation as a National Facility managed by CSIRO.}) to monitor
PMN~J1838--3427, a two-image lensed quasar discovered by Winn et al.\
(2000). The next section summarizes the properties of this object and
the design of our campaign, which was similar to the successful effort
by Lovell et al.\ (1998) to measure the time delay of
PKS~1830--211. Section~\ref{sec:vla} presents new radio maps based on
data from the Very Large Array (VLA\footnote{The VLA is operated by
the National Radio Astronomy Observatory, a facility of the National
Science Foundation operated under cooperative agreement by Associated
Universities, Inc.}). These observations were used to produce a better
model of the radio source structure and thereby improve the analysis
of the ATCA data. The ATCA observations are described in
\S~\ref{sec:atca}, and the data reduction procedure is explained in
\S~\ref{sec:data}. Correlated variability was not detected, and no
time delay could be measured. In fact, excess variability was measured
in the fainter image that cannot easily be explained as intrinsic
variations or systematic errors. In \S~\ref{sec:discussion} we discuss
these possibilities and argue that refractive scintillation is the
best explanation. In \S~\ref{sec:conclusions} we summarize our
observations and conclusions.

\section{Design of the campaign}
\label{sec:design}

The properties of the target object, PMN~J1838--3427, were described
in detail by Winn et al.\ (2000). Here we provide a summary. At
centimeter wavelengths, the system has a flat spectrum and a total
flux density of about 0.3~Jy. It consists of two images (A and B) of a
quasar at a redshift of $z=2.78$, produced by the lensing effect of a
radio-quiet foreground galaxy at $z\approx 0.35$.  Both images are
very compact, even at milliarcsecond resolution.  Image B is
$1\farcs0$ south of image A, and has a flux density that is
approximately 14 times smaller than the flux density of image A.

This object appeared to offer excellent prospects for time-delay
measurement. It is bright, by modern standards, which allows for short
observations at each epoch, and for self-calibration of antenna-based
errors. The flat radio spectrum and the compactness of the radio
components are indicators of radio variability, and indeed, three VLA
and ATCA measurements of the total 8.5~GHz flux density by Winn et
al.\ (2000) ranged from 0.18 to 0.27~Jy. Assuming the galaxy mass
distribution to be isothermal ($\rho \propto r^{-2}$), lens models
predict a time delay of $15h^{-1}$~days.\footnote{Here, $h$ is defined
such that $H_0=100h$~km~s$^{-1}$~Mpc$^{-1}$, and the cosmological
model is assumed to be flat, with $\Omega_{\rm M} = 0.3$ and
$\Omega_\Lambda=0.7$.} This is long enough for a sampling rate of once
every few days to be sufficient, and short enough for the delay to be
measurable in a single season of a few months.

Given its southerly declination, the target is easily observed over a
wide range of hour angles from the ATCA site. Although the ATCA is a
linear array, and therefore provides only one-dimensional spatial
information in a short observation, all that is needed is for some of
the projected baselines to be long enough in the north--south
direction to resolve the $1\arcsec$ double. At an observing frequency
of 9~GHz (the highest frequency that was routinely available at the
time of our campaign), the rule of thumb $\lambda/2D < 1\arcsec$ leads
to the criterion $D > 3.4$~km. All of the ATCA's standard
configurations have at least a few baselines satisfying this criterion
(although the more extended configurations are obviously
preferable). This prevented the campaign from being interrupted by
configuration changes.

Because of the north-south orientation of the double, we did not want
to observe the target near transit, when the projected baselines are
primarily oriented in the east--west direction. Neither did we want to
observe the target close to the horizon, where the projected baselines
are foreshortened, and where shadowing, spillover, and gain-elevation
effects are greatest. Our simulations showed that a good compromise
was to observe the object at an hour angle of about +4.5~hours or
$-4.5$~hours.

These considerations determined our observing strategy. Ideally we
would observe at 9~GHz, with a sampling interval of 2~days (about
one-tenth of the expected time delay), for an entire observing season
of 4 months. Each observation was scheduled as close as possible to
the optimal hour angles mentioned above. The 9~GHz receivers produce a
root-mean-squared (RMS) noise level of 0.12~mJy~beam$^{-1}$ in a 30
minute observation, which is low enough for the thermal noise to be
$<$1\% of the flux density of image B. The consistency in the flux
density scale from epoch to epoch is routinely better than 1\%, and
can be checked via observations of a secondary flux density
calibration source. Our simulations suggested that even in the most
compact configurations, the flux densities of B and A would be
measurable with 2\% accuracy, despite the covariance between the
components (as discussed further in \S~\ref{sec:simulations}). Thus,
if the quasar underwent intrinsic variations of 5-10\% or greater, the
time delay could be measured.

\section{Supporting VLA observations}
\label{sec:vla}

A short observation with a linear array such as the ATCA provides
sparse and nearly one-dimensional coverage of the spatial Fourier
plane (also known as the visibility plane, or $(u,v)$-plane). The
two-dimensional radio source structure cannot be determined
independently from such data. In previous observations, the source
appeared to consist entirely of two point sources, but in the course
of our analysis, we came to suspect that this model was not accurate
enough. Furthermore, measurements with a linear array are particularly
susceptible to confusion with neighboring compact radio
sources. Although there is no source brighter than 50~mJy within a
20$\arcmin$ search radius of the 1.4~GHz NRAO VLA Sky Survey (NVSS;
Condon et al.\ 1998), and the nearest NVSS source is 5$\arcmin$ away
and has a flux density of only 3~mJy, it was possible that a compact
inverted-spectrum source could be contaminating our 9~GHz
visibilities. For these reasons we arranged for VLA observations (with
more complete $(u,v)$-coverage) in order to search for
low-surface-brightness sources and confusing sources that might have
been missed in previous observations.

We obtained snapshots with the VLA in three of its four
configurations: the A array, on 2002~March~4; the B array, on
2002~August~12; and the C array, on 2003~January~5. In all cases, the
total observing bandwidth was 100~MHz per polarization in each of two
frequency bands, centered on 8435~MHz and 8485~MHz. A short
observation of 3C~286 was used to set the flux density scale. The
phase calibration source PMN~J1820--2528 was observed before and after
each observation of PMN~J1838--3427.  The data were calibrated using
standard procedures in AIPS \footnote{The Astronomical Image
Processing System (AIPS) is developed and distributed by the National
Radio Astronomy Observatory.} and mapped with Difmap (Shepherd 1997).

\bigskip

\centerline{~\psfig{file=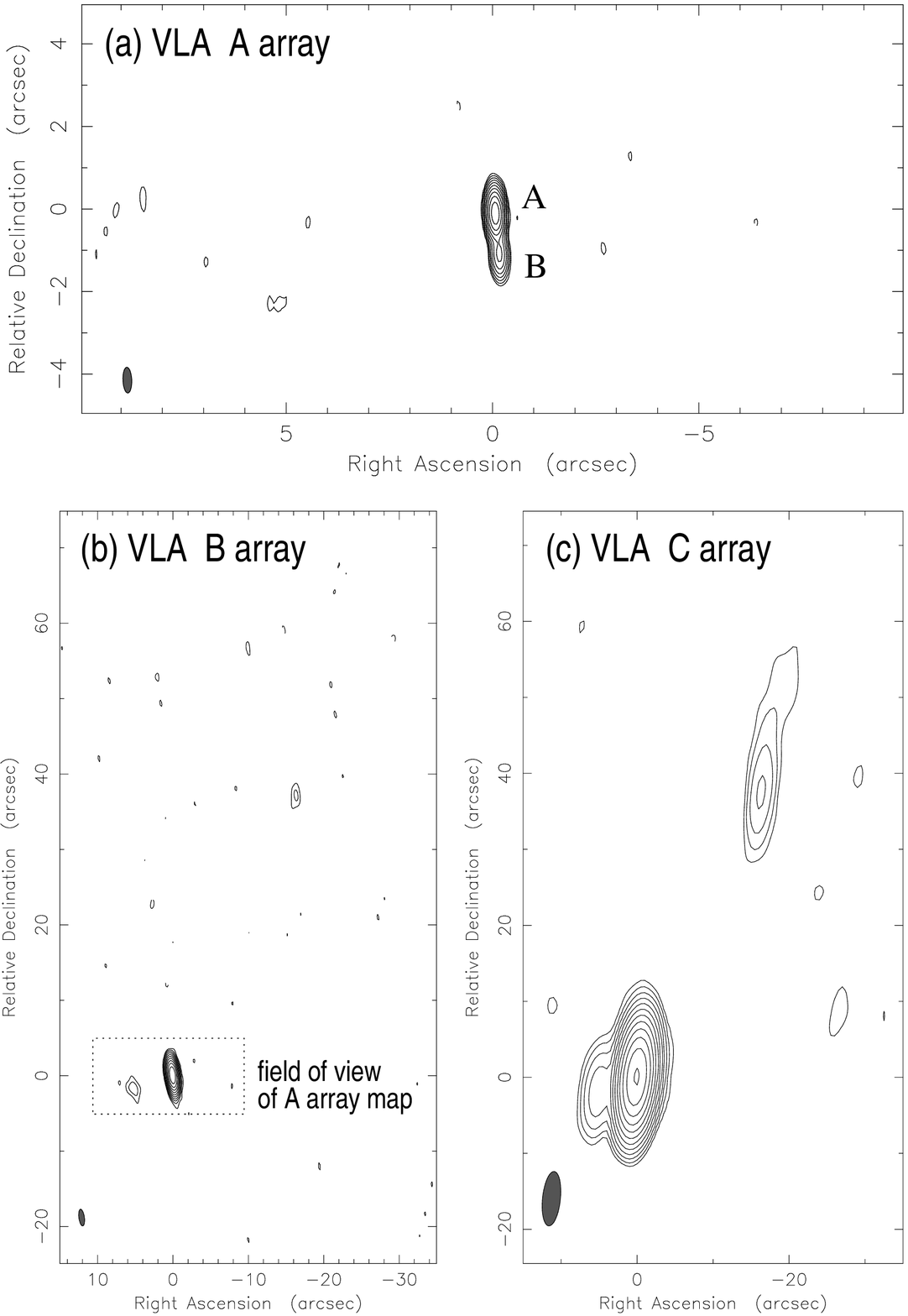,width=3.5in}~}
\figcaption[h]{\footnotesize 8.5~GHz VLA maps of PMN~J1838--3427. All
the maps were created using natural weighting of the visibilities. The
contour levels are $-3\sigma$, $3\sigma$, $6\sigma$, $12\sigma$,
etc. The small gray ellipse in the lower left corner illustrates the
restoring beam.  {\bf (a)} A array map, with FWHM restoring beam
diameters of $0\farcs62\times 0\farcs21$,
$\sigma=0.09$~mJy~beam$^{-1}$. {\bf (b)} B array map, FWHM
$2\farcs2\times 0\farcs7$, $\sigma=0.08$~mJy~beam$^{-1}$. {\bf (c)} C
array map, FWHM $7\farcs2\times 2\farcs4$,
$\sigma=0.03$~mJy~beam$^{-1}$. }
\label{fig1}

\bigskip

Of the three maps, the A array map (Figure~1a) has the highest angular
resolution and the poorest surface-brightness sensitivity. The quasar
images A and B are consistent with point sources. The relative
positions of A and B were measured by fitting a 2-point model to the
visibility data. The results ($\Delta$R.A.~$=97$~mas,
$\Delta\delta=991$~mas) are consistent with the more precise results
obtained by Winn et al.\ (2000) with the Very Long Baseline Array
(VLBA). The B and C array maps (Figures 1b and 1c) reveal two
additional components with lower surface brightness. One of these
components is located about $5\arcsec$ east of B, and has a total flux
density of approximately 1.5~mJy. The other component is located about
$40\arcsec$ to the northwest, and has a total flux density of
approximately 2~mJy. These previously unknown diffuse components are
likely to be radio jets associated with the background quasar,
although it is also possible that one or both components belong to the
foreground object or are unrelated objects.  No other radio sources
were detected in any of the maps.  The widest field of view is
available in the C array map, in which there are no confusing sources
above 0.3~mJy within a radius of $3\arcmin$ (the approximate
half-power point of the primary beam).

\section{ATCA observations}
\label{sec:atca}

The ATCA campaign comprised 54 epochs between 2001~August~31 and
2001~December~31. The average spacing between epochs was 2.4~days. At
each epoch, our observing block was at least one hour in duration,
although some blocks were longer. The observing bandwidth was 128~MHz
per polarization in each of two frequency bands, centered on 8640~MHz
and 8896~MHz. Each band was subdivided into 16 channels of width
8~MHz. All linear polarization products were correlated. Typically we
observed PMN~J1838--3427 for 45 minutes during each session. We also
made short observations of three calibration sources: PKS~1934--638,
the ATCA primary flux density calibration source; PKS~1718--649, a
secondary flux density standard; and PKS~1921--293, a bright and
compact radio source. The latter was intended as a gain calibration
source, although PKS~1718--649 proved to be better for this purpose,
as described below.

The initial calibration was performed with ATCA-specific routines in
the MIRIAD software package (Sault, Teuben, \& Wright 1995). As a
first step, the data were corrected for the elevation-dependence of
the antenna gains, using the task ELEVCOR. This task modifies the
visibilities using gain curves that were empirically determined by
Bignall (2003) for each of the six ATCA antennas. This step is
important for any flux density monitoring program, because the gains
vary by as much as 4\% from the zenith to the lower elevation limit of
12\arcdeg. It is noteworthy that the gain curve for antenna \#6 is
quite different from the gain curves of the other antennas. This means
that the gain--elevation correction is a function of baseline, rather
than being a single multiplicative factor affecting all
visibilities. This is of particular significance for our data because
antenna \#6 forms the longest baselines in the array, and is of
crucial importance in resolving the two components of PMN~J1838--3427.

After the gain--elevation correction, the data were inspected visually
to remove obviously corrupted points and shadowed baselines. The scans
on PKS~1934--638 were used to calibrate the delays, band pass, flux
density scale, and polarization leakage terms.  The scans on
PKS~1718--649 were used to determine the antenna gains.  Because
PKS~1718--649 was observed at nearly the same elevation as
PMN~J1838--3427, this step acted to eliminate residual gain--elevation
effects.  (The other calibration source, PKS~1921--293, is closer on
the sky to the lens, but was typically observed at a significantly
different elevation angle, and provided poorer correction of
gain--elevation effects.)  Data from 11 epochs were lost or unusable
due to problems such as hardware or software glitches, solar
interference, and observer error. The analysis of the data from the
useful 43 epochs is presented below.

\begin{figure*}[t]
\begin{center}
  \leavevmode
\hbox{%
  \epsfxsize=6in
  \epsffile{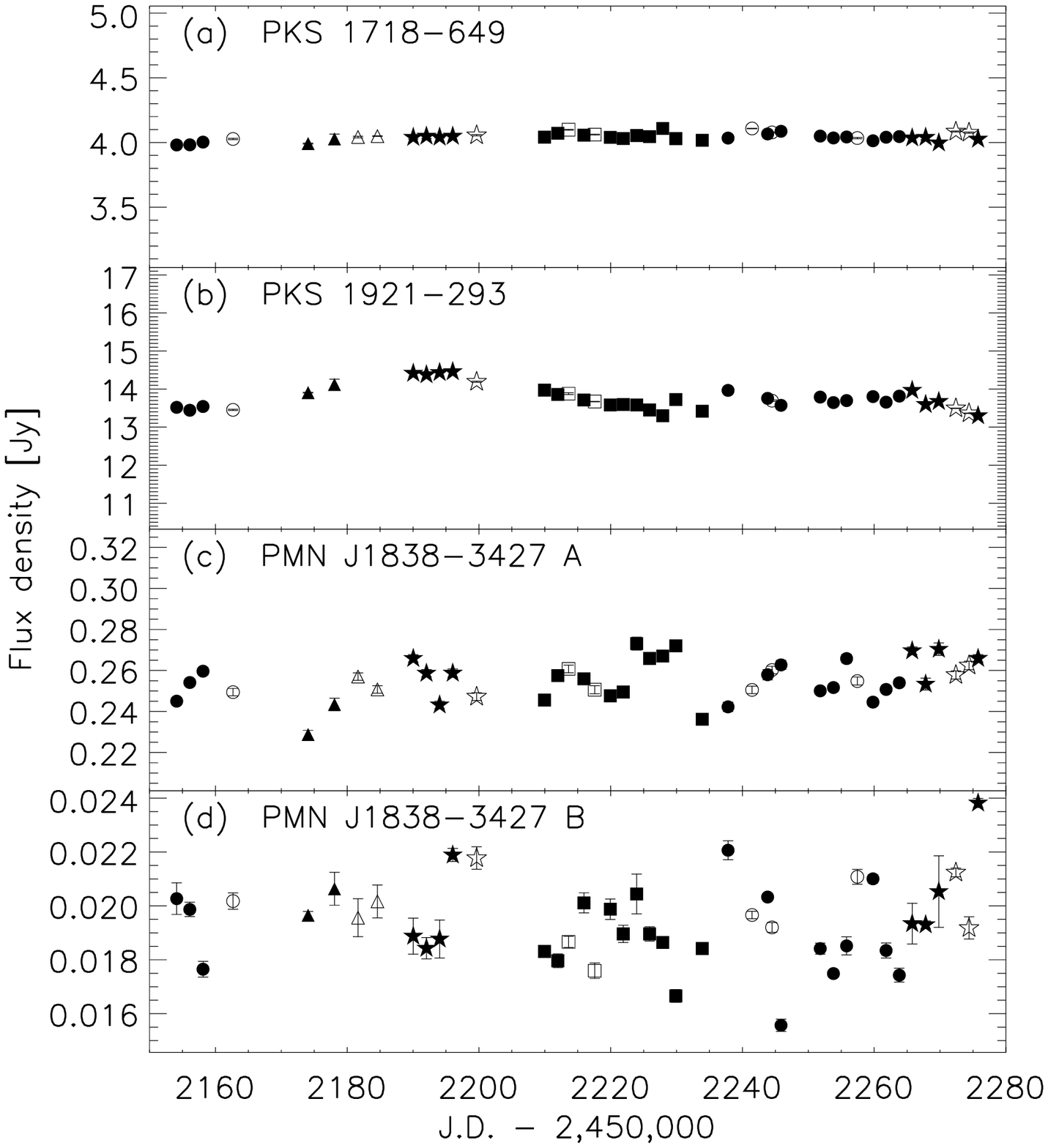}}
\end{center}
\caption{\label{fig2}
ATCA light curves. In each case, the limits of the flux density scale
have been set to $\pm 25$\% of the mean flux density.  The symbol
shapes encode the ATCA array configuration: circles for 6A, 6B, and
6D; squares for 1.5D; triangles for 0.75D; and stars for EW352. Filled
symbols are for measurements at LST$\approx$23 (``late'') and open
symbols are for LST$\approx$14 (``early''). Where error bars are not
seen, they are smaller than the symbol size.
}
\end{figure*}

\section{Data reduction}
\label{sec:data}

\subsection{Light curve of PKS~1718--649}
\label{sec:pks1718}

Gigahertz-peaked spectrum (GPS) sources are often used as
interferometric flux density calibration sources because they tend to
be compact and flux-stable at radio wavelengths (see, e.g., O'Dea
1998). The radio source PKS~1718--649 is a GPS source and is
unresolved by the ATCA at 9~GHz (Tingay et al.\ 1997). These
properties made PKS~1718--649 a potentially useful source for checking
the consistency of the flux density scale.

For each epoch in our campaign, a model consisting of a single point
source was fitted to the visibility data for PKS~1718--649. The
antenna phases were self-calibrated with a 10~s solution interval, and
the model-fitting procedure was repeated. The resulting flux density
was recorded after three more iterations of self-calibration and model
fitting.  The two frequency bands were processed separately until this
point. The mean of the 8896~MHz time series was 1.4\% smaller than the
mean of the 8640~MHz time series, as expected from the previously
measured radio spectrum of the source (Tingay et al.\ 1997). We
increased the 8896~MHz flux densities by 1.4\%, and then averaged the
results of the two bands at each epoch. The final light curve is shown
in Figure~2a. The error bars span the difference between the results
of the two bands (after having normalized the 8896~MHz data). These
error bars give an indication of the statistical noise in each
measurement. The fluctuations of PKS~1718--649 are small, with an RMS
variation of 29~mJy, or 0.7\% of the mean flux density of 4.04~Jy.

The local sidereal times of the epochs were strongly clustered around
two values, 14:00 and 23:00 (hereafter, ``early'' and ``late''),
because of the preference for particular hour angles mentioned in
\S~\ref{sec:design}. In Figure~2a, the open symbols represent data
from the early LSTs, and the filled symbols represent data from the
late LSTs. The different symbol shapes represent different array
configurations.  The reason for encoding this information in the light
curves is that many systematic effects, such as confusion,
gain--elevation effects, or errors in the flux density scale, would
cause correlations of the measured flux density with either the LST or
the array configuration.  In this case, there are no significant
correlations with either LST or array configuration.  (By contrast,
when the data were first analyzed without applying any gain--elevation
corrections, i.e.\ omitting ELEVCOR in the procedure described in
\S~\ref{sec:atca}, there was a strong LST-dependence.)

We conclude that the flux density scale is stable to within 0.7\%, and
may be even better than that, because some of the variations in
PKS~1718--649 could be real variations rather than systematic
errors. Subsequent to our campaign, we learned that variability of
this source has been observed in other monitoring programs [see, e.g.,
Gaensler \& Hunstead (2000), Kedziora-Chudczer et al.\ (2001), Tingay
et al.\ (2003), Tingay \& de Kool (2003)].

\newpage

\subsection{Light curve of PKS~1921--293}
\label{sec:pks1921}

The bright and compact radio source PKS~1921--293 is a blazar that is
known to be variable at radio wavelengths (see, e.g., Dent \& Balonek
1980; Romero, Benaglia, \& Combi 1995; Gaensler \& Hunstead 2000). It
was observed as a potential gain calibrator, although in the end
PKS~1718--649 proved to be more useful for this purpose. Nevertheless
we produced a light curve for PKS~1921--293 in exactly the same way as
we had for PKS~1718--649. The result is shown in Figure~2b. The RMS
variation is 0.3~Jy, or 2.2\% of the mean flux density of 13.8~Jy,
with a time scale of approximately 20~days from peak to peak. In
\S~\ref{sec:scintillation} we consider the possibility that this
variation is due to refractive scintillation, which is relevant to our
interpretation of the light curves of PMN~J1838--3427.

\subsection{Light curves of PMN~J1838--3427 A and B}
\label{sec:pmnj1838}

The flux densities of images A and B ($S_{\rm A}$ and $S_{\rm B}$)
were measured from the ATCA data using a procedure similar to the one
employed for PKS~1718--649 and PKS~1921--293, except that the source
model was not a point source. Instead, the model consisted of two
point sources with relative positions that were fixed at the values
measured in the VLA A~array data, and also a model of the two diffuse
components discovered in the VLA B~array and C~array data
(\S~\ref{sec:vla}). Ideally, the model should be based on a sampling
of the $(u,v)$ plane that is similar to the sampling of the ATCA
monitoring data. The projected baselines of the ATCA data ranged in
length from 0 to 140~k$\lambda$, with most of the data between 0 and
100~k$\lambda$. This matches the range of projected baselines in the B
array data better than the A or C array data. Thus, to describe the
diffuse components, we used a CLEAN-component representation based
upon the B array visibilities with projected baselines shorter than
120~k$\lambda$. The positions and flux densities of the CLEAN
components were held constant during the fits.

The results from the two frequency bands were processed separately and
then averaged, as in \S~\ref{sec:pks1718}. The final light curves are
plotted in Figure~2c and 2d.  The error bars show the quadrature sum
of two quantities: the difference between the results of each
frequency band (as above), and 0.7\% (the uncertainty in the flux
density scale). For all the light curves in Figure~2, the limits on
the flux density axis were chosen to be $\pm 25$\% of the mean flux
density. This allows for an easy comparison of the fractional
variations in each light curve. The hour angle and array configuration
of each observation are encoded in the symbols in Figure~2 (see the
caption).

The RMS variation of $S_{\rm A}$ is 9.8~mJy, or 3.8\% of the mean flux
density of 255~mJy.  The time scale of the variations seems to be
5--10 days from peak to peak, with some outliers.  The amplitude of
the fluctuations is much larger than the statistical error and also
larger than the 0.7\% variation of PKS~1718--649.  We tested for
systematic errors by looking for correlations between $S_{\rm A}$ and
the LST, the array configuration, the flux density of PKS~1718--649,
and with $S_{\rm B}$, but we did not find any significant
correlations. For these reasons, it appears that the variations in
$S_{\rm A}$ are not systematic errors but rather are real variations
of the radio source. Whether they are intrinsic (due to changes in the
quasar luminosity) or extrinsic (due to scintillation) will be taken
up in the next section.

The RMS variation of $S_{\rm B}$ is 1.6~mJy, or 8.0\% of the mean flux
density of 19.7~mJy.  The variations do not appear to be well sampled,
implying that the time scale is $\lsim 2$~days.  Again, the amplitude
of the fluctuations is larger than the statistical error and the error
in the flux density scale, and there are no significant correlations
between $S_{\rm B}$ and the LST, the array configuration, the flux
density of PKS~1718--649, or $S_{\rm A}$.

\section{Discussion}
\label{sec:discussion}

\subsection{Absence of correlated intrinsic variability}
\label{sec:no_time_delay}

Lens models generally predict that intrinsic variations of a 2-image
quasar should be seen first in the image that is observed further from
the center of the lens galaxy. In this case, variations of image A are
expected to precede image B; for a model in which the mass
distribution of the lens galaxy is isothermal, the delay is
$15h^{-1}$~days (Winn et al.\ 2000). If the observed variability were
due entirely to intrinsic variations of the quasar, then one would
expect $S_{\rm B}(t)$ to resemble $S_{\rm A}(t)$ after scaling it by
the magnification ratio and shifting it in time by 15--30
days. Unfortunately, no such correspondence can be discerned in the
light curves of Figure~2. There are significant variations in the
light curve of image A, as argued in the previous section, but they
cannot be matched up to variations in the light curve of image
B. Consequently, the time delay could not be measured from these data.

Although apparent to the eye, this fact was confirmed in the following
manner. For a given trial value of $\tau$, the A--B time delay, we
computed $\sigma(\tau)$, which is defined as the RMS variation of the
``difference light curve''
\begin{equation}
\Delta s(t) = \frac{S_{\rm A}(t-\tau)}{\langle S_{\rm A}\rangle} -
              \frac{S_{\rm B}(t)}     {\langle S_{\rm B}\rangle}.
\end{equation}
Here, $\langle S_{\rm A}\rangle$ and $\langle S_{\rm B}\rangle$ are
the mean values of $S_{\rm A}(t)$ and $S_{\rm B}(t)$. The RMS was
computed over the time range for which there are overlapping data for
$S_{\rm A}(t-\tau)$ and $S_{\rm B}(t)$. The interpolation that is
necessary for computing $S_{\rm A}(t-\tau)$ was performed with cubic
splines. If the variations in the light curves were purely intrinsic,
then $\sigma(\tau)$ would be minimized when $\tau$ is the true time
delay.  The results are shown in Figure~4. There is no clear
minimum. The smallest values of $\sigma$ are obtained for $\tau<0$,
even though we expect $\tau>0$ because variations of image A should
precede variations in image B.  We also note there is no significant
minimum at $\tau=0$, confirming our earlier statement that $S_{\rm
A}(t)$ and $S_{\rm B}(t)$ are not correlated.

\centerline{~\psfig{file=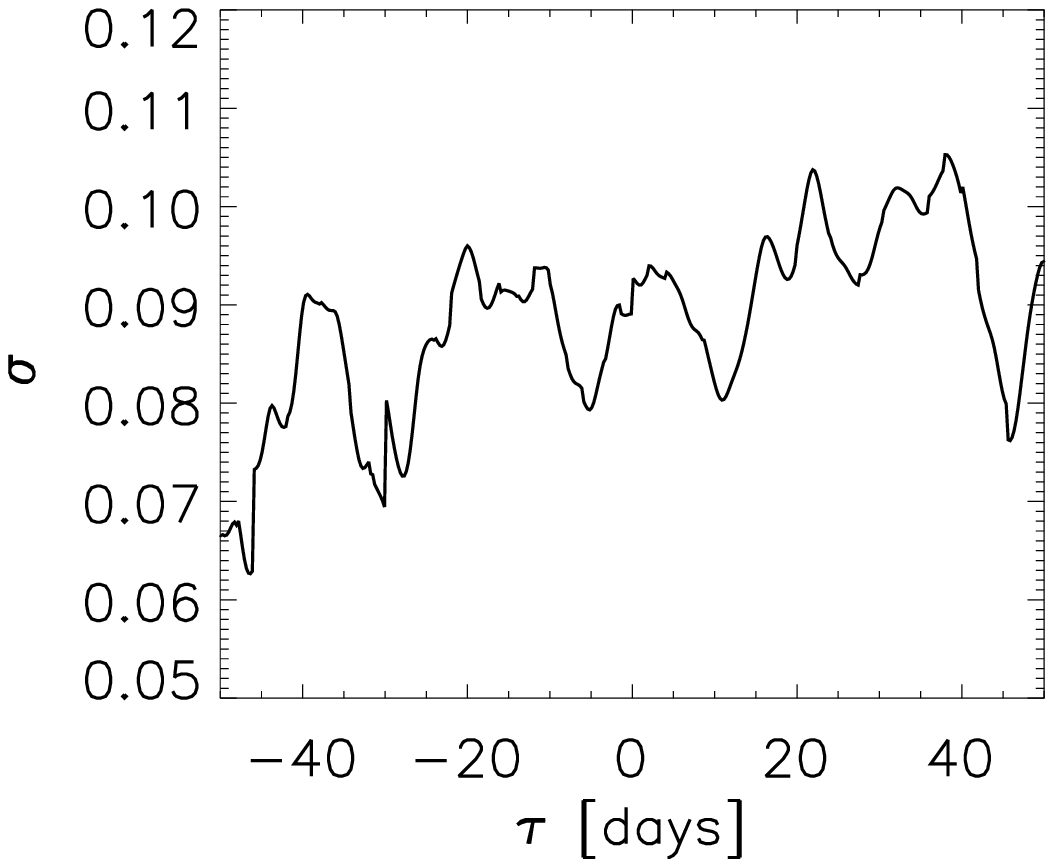,width=3.5in}~}
\figcaption[h]{\footnotesize The RMS variation of the difference light
curve, as a function of the trial value of the time delay. This
quantity should be minimized at the true value of the time delay.  No
clear minimum can be discerned, and therefore no time delay can be
measured. }
\label{fig3}
\bigskip

In the rest of this section, we discuss our attempts to understand the
reason for the lack of correspondence between the light curves. Apart
from the statistical noise (which is negligible for $S_{\rm A}$, and
small even for $S_{\rm B}$), there are three possible sources of
variability in the light curves: intrinsic variations of the radio
source, unaccounted-for systematic errors in the flux density
measurements, and extrinsic variations of the radio source (e.g.\
scintillation). Below we consider these possibilities in turn.

\begin{figure*}[t]
\begin{center}
  \leavevmode
\hbox{%
  \epsfxsize=6in
  \epsffile{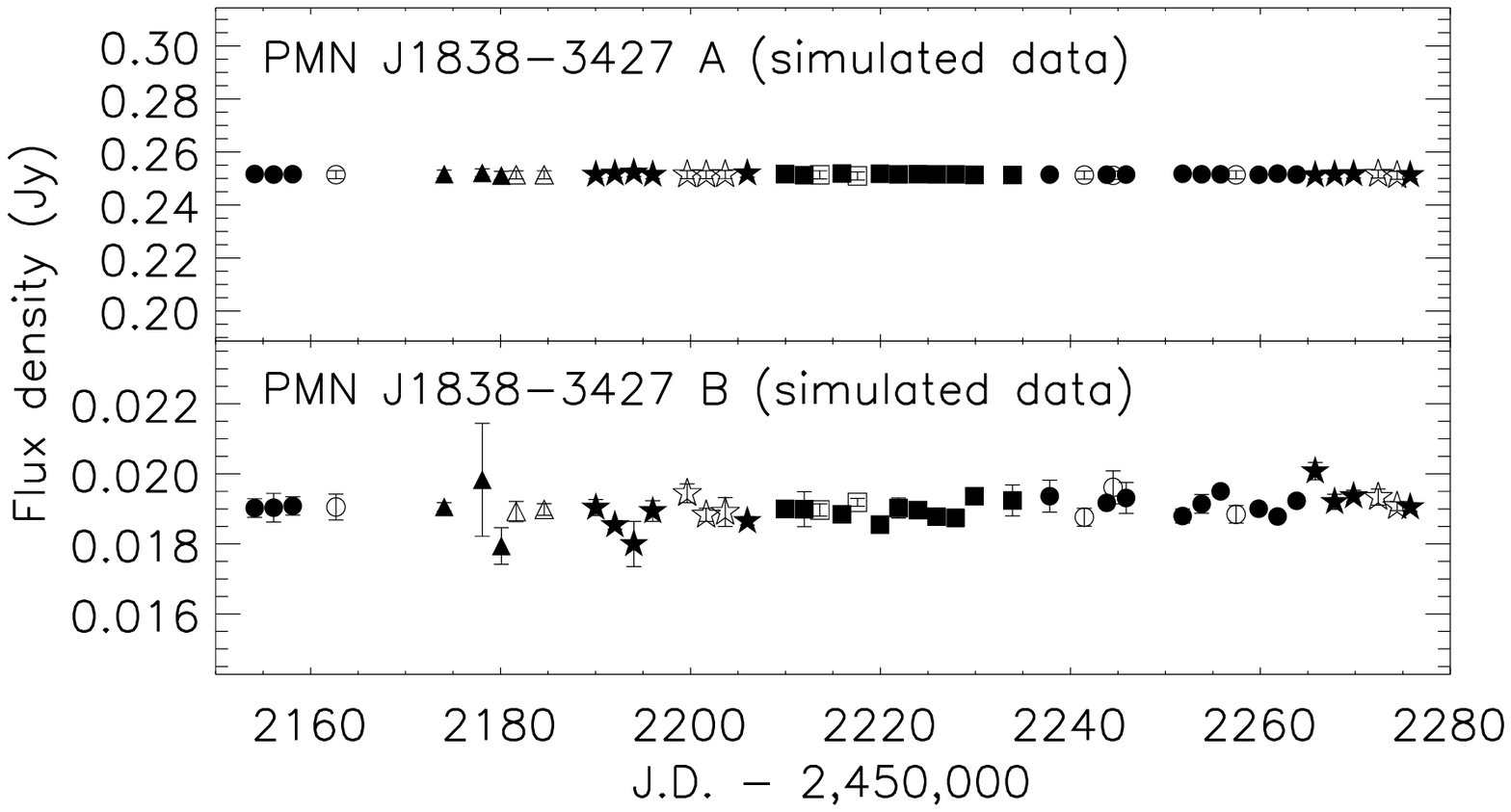}}
\end{center}
\caption{\label{fig4}
ATCA light curves of two images of PMN~J1838--3427, based on
simulated data with the same receiver noise and $(u,v)$-coverage as
the real data. The symbol conventions are the same as in Figure~2.
}
\end{figure*}

\subsection{Excess variability of image B}
\label{sec:excess_rms}

It is unlikely that the observed variability in $S_{\rm B}$ is due
entirely (or even mainly) to intrinsic variations of the source---not
only because no correlated variability is seen, but also because the
fractional variability in $S_{\rm B}$ is larger than that of $S_{\rm
A}$. Suppose the quasar is a point source with intrinsic flux density
$S(t)$. Then one would expect $S_{\rm A}(t) = \mu_{\rm A} S(t-t_0)$
and $S_{\rm B}(t) = \mu_{\rm B} S(t-t_0-\tau)$, where $\mu_{\rm A}$
and $\mu_{\rm B}$ are the image magnifications, $t_0$ is the
light-travel time for image A, and $\tau$ is the time delay. The
fractional variations would be equal, $\frac{\delta S_{\rm B}}{S_{\rm
B}} = \frac{\delta S_{\rm A}}{S_{\rm A}} = \frac{\delta S}{S}$,
whereas in reality $\frac{\delta S_{\rm B}}{S_{\rm B}} \approx
2\frac{\delta S_{\rm A}}{S_{\rm A}}$.  Since the monitoring duration
was 5 times longer than the expected time delay, the excess
variability in image B is unlikely to be a fluke.

One way to escape the conclusion that the variability cannot be
intrinsic is to suppose that the variable portion of image A is
blended with a non-variable component that is not doubly imaged. For
example, image A may have a jet with flux density $S_0$ that has no
counterpart in image B. This would decrease the fractional variability
in $S_{\rm A}$ without affecting the fractional variability in $S_{\rm
B}$. The problem is that one would need $S_0\approx 130$~mJy in order
to dilute the fractional variability of A by a factor of 2, and
although there is evidence for an extended component near image A, it
appears to have a smaller flux density. Winn et al.\ (2000) found the
5~GHz VLBA flux density of image A to be 55~mJy smaller than the VLA
flux density measured 4 months previously. If not due to source
variability, the discrepancy could be caused by an extended portion of
image A that is blended with the compact point source at arcsecond
resolution, but that is nearly invisible at milliarcsecond resolution
(due to the lack of short baselines). Since an unknown fraction of
this extended component is doubly imaged, and since radio jets tend to
have steep radio spectra (i.e.\ the flux density decreases with
increasing frequency), the 9~GHz diluting component should have $S_0 <
55$~mJy.

\subsection{Simulated data}
\label{sec:simulations}

In previous sections we discussed tests for possible systematic errors
due to an inaccurate source model, confusing sources, and residual
gain--elevation effects.  The excess variability of component B does
not seem to be caused by any of these effects.  The light curves did
not change appreciably when the source model was altered to include
the diffuse components discovered with the VLA.  No confusing sources
were found, and there are no detectable phase slopes in the visibility
data that would indicate a confusing source.  There are no significant
correlations between the flux densities and the elevation angle, or
array configuration.

In this section we discuss an additional concern: given that the
longest ATCA baselines were only just sufficient to resolve the
double, there is a potentially large covariance between $S_{\rm A}$
and $S_{\rm B}$. For this reason, as mentioned in \S~\ref{sec:design},
we performed simulations prior to the campaign to check the accuracy
with which we would be able to separate the flux densities of the
components. These simulations suggested we could achieve 2\% accuracy
in $S_{\rm B}$ despite the small A--B separation. Here we discuss a
more detailed simulation based on the actual noise properties of the
data, which confirmed our previous estimate of the achievable
accuracy.

The light curves are based on the Stokes $I$ (total intensity) data,
but the full polarization information was recorded. Since the
fractional circular polarization of PMN~J1838--3427 is smaller than
0.25\%, the Stokes $V$ visibilities from each epoch were nearly
consistent with zero, and they had exactly the same statistical noise
level and $(u,v)$ coverage as the $I$ data. This made the $V$ data
useful for testing purposes.

To the Stokes $V$ visibilities from each epoch, we added a model
visibility function appropriate for PMN~J1838--3427, using the MIRIAD
task UVMODEL. The relative positions of the 2 point sources in the
model were fixed at the values measured with VLBI, and the flux
densities were set at 0.252~Jy and 0.019~Jy. We then produced light
curves from this artificial data set using the same procedure as we
had used on the real data. The resulting light curves are in
Figure~3. The flux density of image A is recovered with 0.1\%
precision. There are fluctuations in $S_{\rm B}$ due to the noise in
the visibilities, the model fitting procedure, and varying
$(u,v)$-coverage, but the RMS fractional variation is only 2\%. The
8\% variations observed in the actual data must have a different
origin.

\subsection{Extrinsic variability}
\label{sec:scintillation}

Compact radio sources scintillate due to scattering by the ionized
interstellar medium (ISM), much as stars twinkle due to scattering by
the Earth's atmosphere (see, e.g., Rickett 1990 or Narayan 1992). Is
scintillation a plausible explanation for the lack of correlation
between the light curves, and in particular for the excess variability
of image B? In this section we ask what type and degree of
scintillation one might expect to occur along this line of sight
through the Galaxy, and whether the amplitude, bandwidth, and time
scale of the observed variations are consistent with those
expectations.

Walker (1998, 2001) used the Taylor \& Cordes (1993) model of the
ionized ISM to give rough expectations for the scattering of
extragalactic radio sources. According to this model, for the line of
sight to PMN~J1838--3427 (Galactic longitude $l_{II}=0\fdg4$ and
latitude $b_{II}=-12\fdg5$), the transition frequency between strong
(multi-path) and weak scattering is $\nu_0 \approx 30$~GHz, with
strong scattering occurring for $\nu<\nu_0$ as in our
observations. Strong scattering phenomena can be divided into
diffractive scintillation, which refers to the interference pattern
produced by the multiple image paths, and refractive scintillation,
which refers to the focusing and de-focusing of the entire image
ensemble (the ``scatter-broadened'' image). Diffractive scintillation
requires a source with an exceptionally small angular size or an
exceptionally nearby scattering screen, and is rarely observed for
extragalactic radio sources. Hence for our purposes the relevant
phenomenon is probably refractive scintillation.

Refractive scintillation produces an RMS intensity modulation of
\begin{equation}
m = \left(\frac{\nu}{\nu_0}\right)^{17/30}
    \left(\frac{\theta_{\rm R}}{\theta_{\rm S}}\right),
\end{equation}
where $\theta_{\rm S}$ is the angular size of the source and
$\theta_{\rm R}$ is the size of the refractive scattering disk. Using
the estimates of Walker (2001) we obtain $\theta_{\rm R}\approx
15$~$\mu$as for the line of sight to PMN~J1838--3427 at 9~GHz. For
scintillation to produce the observed 4\% modulation of image A, a
source size of $\theta_{\rm S}\approx 0.2$~mas is needed. This is
consistent with being unresolved in the 5~GHz VLBA map of Winn et al.\
(2000), which had an angular resolution of $10.5\times 1.8$~mas. It is
also consistent with the theoretical lower limit on angular size due
to the well known ``inverse Compton catastrophe'' (Kellermann \&
Pauliny-Toth 1969): $\theta_{\rm S} > (0.6$~mas)~$\nu^{-1}\sqrt{S}$,
where $\nu$ is in GHz and $S$ is in Janskys. In this case, the
background radio source has a flux density of $S\approx 0.1$~Jy,
corresponding to a minimum angular size of 20~$\mu$as. The angular
size of image A is approximately twice the intrinsic source size, due
to lensing magnification, giving $\theta_{\rm A} > 40$~$\mu$as. Thus,
both the observed upper limit and theoretical lower limit on the
angular size of image A are consistent with the scintillation
hypothesis. Furthermore, one naturally expects the fractional RMS
variation of image B to be a few times larger than that of image A,
because the angular size of image B is a few times smaller.\footnote{A
caveat is that the line of sight to image B passes much closer to the
center of the lens galaxy than that of image A. This raises the
possibility that the angular size of image B may be larger than
expected due to scatter-broadening by the ISM of the lens galaxy, as
has been observed in PKS~1830--211 by Jones et al.\ (1996) and in
PMN~J0134--0931 by Winn et al.\ (2003).}

Refractive scintillation is by nature a broad band phenomenon
($\Delta\nu \sim \nu$), which is consistent with our observation that
the modulations were the same in two frequency bands separated by
$\Delta\nu = 0.03\nu$. The time scale of the variations depends upon
the transverse speed of the scattering screen. Walker (1998), assuming
$v\approx 50$~km~s$^{-1}$, found the time scale to be
\begin{equation}
t\approx (1.4 \hskip 0.025in {\rm hr}) \hskip 0.04in
        \left(\frac{\nu_0}{\nu}\right)^{11/5}
        \left(\frac{\theta_S}{\theta_R}\right).
\end{equation}
This gives approximately 10 days for $\theta_S=0.2$~mas, in agreement
with the light curve for image A. The time scale is expected to be a
few times shorter for image B, again because the angular size of image
B is smaller than that of image A. This is also consistent with the
light curves. The lack of correlation between the light curves is also
expected, because the 15~$\mu$as scattering disk is much smaller than
the $1\arcsec$ separation between the two images. In short, refractive
scintillation is expected for this source at 9~GHz, and it can account
for the observed amplitudes and time scales of variability of both
components.

Having put forth this hypothesis for the observed variations of
PMN~J1838--3427, one might wonder whether scintillation should also
have been observed for PKS~1921--293 or PKS~1718--649. For the line of
sight to PKS~1921--293 ($l_{II}=9\fdg3$, $b_{II}=-19\fdg6$), Walker
(2001) estimates $\nu_0\approx 15$~GHz and $\theta_{\rm R}\approx
5.2$~$\mu$as. The 9~GHz source size is $\lsim$0.4~mas, given the VLBA
map of Fey, Clegg, \& Fomalont (1996), and therefore the modulation
index due to scintillation should be $\gsim$0.01 and the time scale of
the variations should be $\lsim$15~days. These figures are in fair
agreement with the observed RMS variations of 2.2\% on a time scale of
about 20 days. Of course, it is possible that some of the observed
variations are intrinsic to the blazar, but at least the scintillation
hypothesis does not predict more variation than is observed, and
therefore does not contradict our observations.

Likewise, for PKS~1718--649 ($l_{II}=327\fdg0$, $b_{II}=-15\fdg8$),
Walker (2001) estimates $\nu_0\approx 15$~GHz and $\theta_{\rm
R}\approx 4.6$~$\mu$as.  Tingay et al.\ (2002) showed that the source
consists of two components, each about 1~mas in diameter, separated by
about 7~mas.  Each component should scintillate with RMS variations of
approximately 0.3\% on a time scale of $\approx$40~days, which is
consistent with our observation of little or no fluctuations in flux
density.

Finally, we note that Koopmans \& de Bruyn (2000) have argued that
there is another potential source of extrinsic variability:
microlensing due to compact masses in the lens galaxy. It is generally
thought that the angular sizes of radio sources are too large to be
significantly affected by microlensing by stellar-mass objects. In
contrast, Koopmans \& de Bruyn (2000) argue that relativistic beaming
can allow for radio microlensing by shrinking the effective source
size, and that this has been observed in the 2-image lens
B1600+434. The basis for their argument that the observed variability
is due to microlensing, rather than scintillation, is the frequency
dependence of the RMS modulation. Since that information is not
available in this case, and since the more conventional explanation of
scintillation appears entirely plausible, we do not discuss
microlensing further.

\section{Conclusions}
\label{sec:conclusions}

In an effort to measure the time delay of the two-image gravitational
lens PMN~J1838--3427, we monitored the system at 9~GHz for 4 months
with the ATCA. We achieved a stability in the flux density scale of
0.7\% or better. The brighter quasar image varied at the 4\% level,
and the fainter quasar image varied at the 8\% level.  It appears
likely that scintillation by the ionized ISM caused the observed
variability. The time delay could not be measured from these data
because no intrinsic correlated variations could be identified.
Although this campaign was not successful in this regard, it is our
hope that the description of our campaign, data analysis, and
interpretation will be of use in future interferometric monitoring
campaigns.

The scintillation hypothesis can be tested further with
multi-frequency monitoring, by comparing the frequency dependence of
the observed variability with the expectations of scattering theory
(with the caveat that the scattering medium may not be as simple as
the idealized Kolmogorov medium that is assumed in such calculations).
Future 9~GHz VLBI maps of PMN~J1838--3427 may resolve image A at the
$\sim$0.2~mas level. Longer term monitoring might reveal the annual
variations in time scale that are sometimes observed in scintillating
sources (for a recent example, see Bignall et al.\ 2003). Apart from
merely explaining the variability of PMN~J1838--3427 observed in our
campaign, the observation of two such closely spaced scintillating
components may be of intrinsic interest. The time scales and
amplitudes of the variations will depend upon the relative dimensions
of the two images along the direction of motion of the scattering
screen. These inferred dimensions could be compared with the
gravitational lensing model to estimate the direction of motion of the
scattering screen. Unknowns such as the intrinsic source morphology,
the degree of anisotropic scattering, and possible broadening of the
images due to scattering in the lens galaxy, would confuse the
issue. Perhaps if one could identify a scintillating four-image lens,
the greater number of constraints would allow one to disentangle some
of these effects.

\acknowledgments We are grateful to Ramesh Narayan for helpful
discussions regarding scintillation, Vince McIntyre and Robin Wark for
assistance with the ATCA observations, and Barry Clark for allocating
{\it ad hoc} VLA time for supporting observations. This work was
supported by the National Science Foundation under Grant No.\ 0104347.

\end{document}